\documentclass[a4paper]{amsart}
\usepackage{epic,eepic,amsmath,amsthm,amssymb}
\usepackage{epsfig,cite,indentfirst,oldgerm}
\usepackage{multicol,boxedminipage}

\begin{document}

\title[Hall-Littlewood and KP ]
{Hall-Littlewood plane partitions and KP}

\author{O Foda and M Wheeler}

\address{Department of Mathematics and Statistics,
         University of Melbourne, 
         Parkville, Victoria 3010, Australia.}
\email{foda, mwheeler@ms.unimelb.edu.au}

\date{}

\newcommand{\field}[1]{\mathbb{#1}}
\newcommand{\C}{\field{C}}
\newcommand{\N}{\field{N}}
\newcommand{\Z}{\field{Z}}
\newcommand{\R}{\field{R}}

\begin{abstract}
MacMahon's classic generating function of random plane 
partitions, which is related to Schur polynomials, was 
recently extended by Vuleti\'c to a generating function 
of weighted plane partitions that is related to 
Hall-Littlewood polynomials, $\mathcal{S}(t)$, and further 
to one related to Macdonald polynomials, $\mathcal{S}(t,q)$.

Using Jing's 1-parameter deformation of charged free fermions, 
we obtain a Fock space derivation of the Hall-Littlewood 
extension. Confining the plane partitions to a finite 
$s\times s$ square base, we show that the resulting
generating function, $\mathcal{S}_{s\times s}(t)$, is 
an evaluation of a $\tau$-function of KP.
\end{abstract}

\maketitle
\newtheorem{ca}{Figure}
\newtheorem{corollary}{Corollary}
\newtheorem{de}{Definition}
\newtheorem{definition}{Definition}
\newtheorem{example}{Example}
\newtheorem{ex}{Example}
\newtheorem{lemma}{Lemma}
\newtheorem{no}{Notation}
\newtheorem{proposition}{Proposition}
\newtheorem{pr}{Proposition}
\newtheorem{remark}{Remark}
\newtheorem{re}{Remark}
\newtheorem{theorem}{Theorem}
\newtheorem{theo}{Theorem}

\def\ll{\left\lgroup}
\def\rr{\right\rgroup}

\newcommand{\Proof}{\medskip\noindent {\it Proof: }}
\def\no{\nonumber}
\def\ni{\noindent}
\def\proofend{\ensuremath{\square}}
\def\pr{'}
\def\beqa{\begin{eqnarray}}
\def\eeqa{\end{eqnarray}}
\def\ba{\begin{array}}
\def\ea{\end{array}}
\def\gl{\begin{swabfamily}gl\end{swabfamily}}
\def\psis{\psi^{*}}
\def\Psis{\Psi^{*}}
\def\union{\mathop{\bigcup}}
\def\vac{|\mbox{vac}\rangle}
\def\cav{\langle\mbox{vac}|}
\def\dprod{\mathop{\prod{\mkern-29.5mu}{\mathbf\longleftarrow}}}
\def\rprod{\mathop{\prod{\mkern-28.0mu}{\mathbf\longrightarrow}}}
\def\r{\rangle}
\def\l{\langle}
\def\a{\alpha}
\def\b{\beta}
\def\hb{\hat\beta}
\def\d{\delta}
\def\g{\gamma}
\def\e{\epsilon}
\def\tg{\operatorname{tg}}
\def\ctg{\operatorname{ctg}}
 \def\sh{\operatorname{sh}}
 \def\ch{\operatorname{ch}}
\def\cth{\operatorname{cth}}
 \def\th{\operatorname{th}}
\def\eps{\varepsilon}
 \def\la{\lambda}
\def\tla{\tilde{\lambda}}
\def\Gh{\widehat{\Gamma}}
\def\tmu{\tilde{\mu}}
\def\s{\sigma}
\def\sul{\sum\limits}
\def\pl{\prod\limits}
\def\lt({\left(}
\def\rt){\right)}
\def\pd #1{\frac{\partial}{\partial #1}}
\def\const{{\rm const}}
\def\argum{\{\mu_j\},\{\la_k\}} 
\def\umarg{\{\la_k\},\{\mu_j\}} 
\def\prodmu #1{\prod\limits_{j #1 k} \sinh(\mu_k-\mu_j)}
\def\prodla #1{\prod\limits_{j #1 k} \sinh(\lambda_k-\lambda_j)}
\newcommand{\bl}[1]{\makebox[#1em]{}}
\def\tr{\operatorname{tr}}
\def\Res{\operatorname{Res}}
\def\det{\operatorname{det}}
\def\ivac{\langle\Omega|}
\def\fvac{|\Omega\rangle}
\def\fdirac{|0\rangle}
\def\idirac{\langle0|}
\def\psis{\psi^{*}}
\def\Psis{\Psi^{*}}
\def\lprod{\mathop{\prod{\mkern-29.5mu}{\mathbf\longleftarrow}}}
\def\rprod{\mathop{\prod{\mkern-28.0mu}{\mathbf\longrightarrow}}}
\def\complex{\mathbb{C}}
\def\integer{\mathbb{Z}}

\newcommand{\boldN}{\boldsymbol{N}}
\newcommand{\bra}[1]{\langle\,#1\,|}
\newcommand{\ket}[1]{|\,#1\,\rangle}
\newcommand{\bracket}[1]{\langle\,#1\,\rangle}
\newcommand{\infinity}{\infty}

\renewcommand{\labelenumi}{\S\theenumi.}

\let\up=\uparrow
\let\down=\downarrow
\let\tend=\rightarrow
\hyphenation{boson-ic
             ferm-ion-ic
             para-ferm-ion-ic
             two-dim-ension-al
             two-dim-ension-al
             rep-resent-ative
             par-tition
	     anti-comm-uta-tion}

\newtheorem{Theorem}{Theorem}[section]
\newtheorem{Corollary}[Theorem]{Corollary}
\newtheorem{Proposition}[Theorem]{Proposition}
\newtheorem{Conjecture}[Theorem]{Conjecture}
\newtheorem{Lemma}[Theorem]{Lemma}
\newtheorem{Example}[Theorem]{Example}
\newtheorem{Note}[Theorem]{Note}
\newtheorem{Definition}[Theorem]{Definition}
                                                                               
\renewcommand{\mod}{\textup{mod}\,}
\newcommand{\wt}{\text{wt}\,}

\newcommand{\T}{{\mathcal T}}
\newcommand{\U}{{\mathcal U}}
\newcommand{\tT}{\tilde{\mathcal T}}
\newcommand{\tU}{\widetilde{\mathcal U}}
\newcommand{\Y}{{\mathcal Y}}
\newcommand{\B}{{\mathcal B}}
\newcommand{\D}{{\mathcal D}}
\newcommand{\M}{{\mathcal M}}
\renewcommand{\P}{{\mathcal P}}

\hyphenation{And-rews
             Gor-don
             boson-ic
             ferm-ion-ic
             para-ferm-ion-ic
             two-dim-ension-al
             two-dim-ension-al}

\setcounter{section}{-1}

\section{Introduction}\label{introduction}

In \cite{v2}, Vuleti\'c obtained 1-parameter and 2-parameter
deformations of MacMahon's generating function of random 
plane partitions. In this work, we limit our attention 
mostly to the 1-parameter result

\begin{equation}
\mathcal{S}(t) = 
\prod_{j=1}^{\infty}
\ll
\frac{1-t z^j}{1-z^j}
\rr^j
\label{def-pp}
\end{equation}

\noindent where $t \in \C$ is the deformation parameter, 
and $z$ is a formal parameter that keeps track of the 
number of boxes in a plane partition (see below). Setting 
$t=0$, one recovers MacMahon's generating function of random 
plane partitions \cite{mac}, which is an evaluation of 
a KP $\tau$-function \cite{mjd}.

The derivation of (\ref{def-pp}) in \cite{v2} was combinatorial. 
The purpose of this work is two-fold. {\bf 1.} To use Jing's 
1-parameter deformation of free fermions \cite{j1}, to obtain 
a Fock space derivation of (\ref{def-pp}). {\bf 2.} To show that 
$\mathcal{S}(t)$ is an evaluation of a KP $\tau$-function. 

In section {\bf {\ref{HL}}}, we recall Jing's 1-parameter 
$t$-deformation of charged free fermions, Heisenberg algebra,
vertex operators, {\it etc.} \cite{j1}. In section {\bf {\ref{P}}}, 
we introduce Young diagrams\footnote{We use \lq Young diagrams\rq\ , 
or \lq diagrams\rq\ , rather than \lq partitions\rq\  to clearly 
distinguish between (regular) partitions and plane partitions.} 
as labels of (left and right) state vectors in (left and right) 
Fock spaces and evaluate their inner products. 

In section {\bf {\ref{I}}}, we show that the action of a certain 
$t$-vertex operator on a state vector labeled by a Young diagram 
$\mu$ generates a weighted sum of state vectors labeled by Young
diagrams
that interlace with $\mu$. The weights are skew Hall-Littlewood 
functions. The derivation is elementary in the sense that it follows 
directly from the properties of the underlying fermions. This 
extends a result of Okounkov and Reshetikhin for undeformed 
fermions and Schur functions \cite{or2}.

In section {\bf {\ref{PP}}} we compute an expectation value
of (infinitely many) vertex operators in two different ways. 
{\bf A.} By acting on Young diagrams to generate weighted interlacing 
Young diagrams that stack to generate weighted plane partitions. 
{\bf B.} By commuting the vertex operators in such a way that 
they act on the vacuum states as the identity and thereby obtain 
a product expression for the expectation value. Equating 
the results of these two computations, we recover the 1-parameter 
result of \cite{v2}.

In section {\bf {\ref{TF}}} we show that $\mathcal{S}(t)$
is an evaluation of a KP $\tau$-function, for arbitrary values of $t$.  
In section {\bf {\ref{M}}}, we comment on the 2-parameter extension 
of MacMahon's generating function that is related to Macdonald 
polynomials \cite{v2}. In this case, a 2-parameter 
$(t, q)$-deformation of charged free fermions is also available 
\cite{j2}. Using the corresponding $(t, q)$-vertex operators, it
is straightforward to compute the corresponding expectation value 
and obtain the analogue of computation {\bf B} above. What is beyond 
the scope of this work is to obtain the analogue of computation 
{\bf A} in terms of the action of underlying fermions. In section 
{\bf {\ref{C}}}, we collect a number of comments. 

\section{$t$-Fermions and related operators}
\label{HL}

\subsection{The deformation parameter $t$} We take 
$t = e^{-\epsilon + i \theta}$, where $\epsilon, \theta \in \mathbb{R}$, 
$\epsilon > 0$. The fact that $ \epsilon> 0$ will be necessary for 
convergence in the intermediate steps of all derivations. However, ultimately 
it will be possible to take the limits $\epsilon \rightarrow 0$, and 
$\theta \rightarrow 2 \pi / n$, $n \in \N \geq 2$. The limit 
$\epsilon \rightarrow 0$ is justified by the fact that the expressions 
obtained are well defined. 

\subsection{Charged $t$-fermions and $t$-anti-commutation
relations}

Consider two species of charged fermions, $\{\psi_m, \psis_m \}$, 
which satisfy the following $t$-deformed anti-commutation relations

\begin{eqnarray}
\psi_m \psi_n + \psi_n \psi_m &=& 
t \psi_{m+1} \psi_{n-1} + t \psi_{n+1} \psi_{m-1}
\label{a-c1}
\\
\psis_m \psis_n + \psis_n \psis_m &=& 
t \psis_{m-1} \psis_{n+1} + t \psis_{n-1} \psis_{m+1}
\label{a-c2}
\\
\psi_m \psis_n + \psis_n \psi_m &=& 
t \psi_{m-1} \psis_{n-1} + t \psis_{n+1} \psi_{m+1} 
+ 
(1-t)^2 \delta_{m,n}
\label{a-c3}
\end{eqnarray}

\noindent where $m, n \in \Z$, and $t \in \complex$ is a deformation 
parameter. We refer to these as $t$-fermions. For $t=0$, we recover 
the charged free fermions of KP theory \cite{mjd}. 

\subsection{Re-writing the $t$-anti-commutation relations}

Given the structure of the $t$-anti-commutation 
relations (\ref{a-c1}--\ref{a-c3}), it will be useful in 
later sections to re-write them differently. 
Using (\ref{a-c3}) on itself, we obtain

\begin{eqnarray}
\nonumber
\psi_m \psis_n
&=&
-\psis_n \psi_m
+
t \psi_{m-1} \psis_{n-1}
+
t \psis_{n+1} \psi_{m+1}
+
(1-t)^2 \delta_{m,n}
\\ 
\nonumber
&=&
-\psis_n \psi_m
+ t \ll
-\psis_{n-1} \psi_{m-1}
+ t \psi_{m-2} \psis_{n-2}
+ t \psis_n \psi_m
+ (1-t)^2 \delta_{m,n} \rr
\\
\nonumber
&+&
t \psis_{n+1} \psi_{m+1}
+
(1-t)^2 \delta_{m,n}
\\
\nonumber
&=&
t\psis_{n+1}\psi_{m+1}
+
(t^2-1) \psis_n \psi_m
-
t \psis_{n-1} \psi_{m-1}
\\
\nonumber
&+&
t^2 \psi_{m-2} \psis_{n-2}
+
(1+t)(1-t)^2 \delta_{m,n}
\\
\nonumber
&\vdots&
\\
&=&
t \psis_{n+1} \psi_{m+1}
+
(t^2-1) \sum_{j=0}^{\infty} \psis_{(n-j)} \psi_{(m-j)} t^j
+
(1-t) \delta_{m,n}
\label{a-c4}
\end{eqnarray}

\noindent which is clearly a re-writing of (\ref{a-c3}) in 
such a way that a $\psis$ operator appears on the left and 
a $\psi$ on the right in each bilinear term on the right 
hand side of the relation. 

Further, we propose the following identity
\begin{equation}
\psis_{m}\psis_{(m-n)} + (1-t)
\sum_{j=1}^{n} \psis_{(m-j)}\psis_{(m+j-n)} = t \psis_{(m-1-n)}
\psis_{(m+1)}, \quad n \geq 0
\label{a-c5}
\end{equation}

We refer to the statement in (\ref{a-c5}) as $\mathcal{P}_n$, and 
prove it inductively as follows. Rearranging terms in $\mathcal{P}_n$, 
and relabelling of the summation parameter $j$, we obtain

\begin{equation*}
\begin{split}
& \psis_{m}\psis_{(m-n)} + (1-t)
\sum_{j=1}^{n} \psis_{(m-j)}\psis_{(m+j-n)}
=
\\
& \psis_{m}\psis_{(m-n)}
-t \psis_{(m-1)}\psis_{(m+1-n)}
+(1-t)\psis_{(m-n)} \psis_{m}
+
\\
& \ll
\psis_{(m-1)}\psis_{(m-1)-(n-2)}
+(1-t)
\sum_{j=1}^{n-2}\psis_{(m-1)-j}\psis_{(m-1)+j-(n-2)}
\rr
\end{split}
\end{equation*}

If $\mathcal{P}_{(n-2)}$ holds, then the parenthesised term is 
equal to $t \psis_{(m-n)} \psis_m$, and 

\begin{equation*}
\begin{split}
\psis_{m} \psis_{(m-n)} +& 
(1-t) \sum_{j=1}^{n} \psis_{(m-j)}\psis_{(m+j-n)} = \\
\psis_{m} \psis_{(m-n)} -& 
t \psis_{(m-1)}\psis_{(m+1-n)} + \psis_{(m-n)} \psis_{m}
= t\psis_{(m-1-n)}\psis_{(m+1)}
\end{split}
\end{equation*}

\noindent where we have used the $t$-anti-commutation relation
(\ref{a-c2}) in the last line. Hence $\mathcal{P}_n$ true if 
$\mathcal{P}_{n-2}$ is true. But $\mathcal{P}_0$ and 
$\mathcal{P}_1$ follow trivially from (\ref{a-c2}). They are

\begin{eqnarray*}
&\mathcal{P}_0:&
\psis_m \psis_m = t \psis_{(m-1)} \psis_{(m+1)}
\label{n=0}
\\
&\mathcal{P}_1:&
\psis_m\psis_{(m-1)} + (1-t) \psis_{(m-1)}\psis_{m} =
t\psis_{(m-2)}\psis_{(m+1)}
\label{n=1}
\end{eqnarray*}

\noindent and $\mathcal{P}_n$ is true for all $n\in\N$. From 
the proof, it is clear that (\ref{a-c5}) is a re-writing of 
(\ref{a-c2}). An analogous result holds for (\ref{a-c1}), 
but we will not need it in the sequel.

\subsection{$t$-Heisenberg operators}

The $t$-analogues of the Heisenberg generators $h_m$ are defined 
in terms of the $t$-fermions as 

\begin{equation}
h_m 
=
\left\{
\begin{array}{cc}
\frac{1}{(1-t)}
\displaystyle{\sum_{j\in \integer}}
\psi_j \psis_{(j+m)}
&
\quad
m \geq 1
\\
\frac{1}{(1-t)(1-t^{|m|})}
\displaystyle{\sum_{j\in \integer}}
\psi_j\psis_{(j+m)}
&
\quad
m \leq -1
\end{array}
\right.
\label{bosons} 
\end{equation}

Using the $t$-anti-commutation relations (\ref{a-c1}-\ref{a-c3}), 
one can show that

\begin{eqnarray}
h_m \psi_j  - \psi_j  h_m &=& \phantom{-}  \psi_{(j-m)}
\label{com1} 
\\
h_m \psis_j - \psis_j h_m &=&          -  \psis_{(j+m)}
\label{com2}
\\
h_m h_n - h_n h_m &=& \frac{m}{1-t^{|m|}} \delta_{m+n,0}
\label{commutator}
\end{eqnarray}

\noindent where $j \in \Z$, and $m, n \in \Z \backslash \{0\}$. 

\subsection{$t$-Vertex operators}

The $t$-analogue of the vertex operators of \cite{or2} are 

\begin{eqnarray}
\Gamma_{+}(z,t) 
&=& 
\exp 
\ll 
- \sum_{m=1}^{\infty} \frac{1-t^m}{m} z^{-m} h_m 
\rr
\\
\Gamma_{-}(z,t) &=& \exp 
\ll 
-\sum_{m=1}^{\infty} \frac{1-t^m}{m} z^m h_{-m} 
\rr
\end{eqnarray}

They satisfy the commutation relations

\begin{equation}
\Gamma_{+}(z,t) \Gamma_{-}(z',t) = 
\frac{z-t z'}{z-z'} \Gamma_{-}(z',t) \Gamma_{+}(z,t)
\label{h-v3}
\end{equation}

\subsection{$t$-Fermion evolution equations} 

Define the generating series $\Psi(k)=\sum_{j\in \mathbb{Z}} \psi_j k^j$, 
$\Psis(k) = \sum_{j\in \mathbb{Z}} \psis_j k^j$, and 
$H_{\pm}(z,t) = \sum_{n \in \pm \mathbb{N}} \frac{t^{|n|}-1}{n z^n} h_n$.
Using the relations $(\ref{com1})$ and $(\ref{com2})$, we obtain

\begin{eqnarray}
[\Psi(k),H_{+}(z,t)] 
&=& 
\Psi(k) \sum_{n \in \mathbb{N}} \frac{1- t^{n}}{n z^n} k^n 
=
\Psi(k) \log \ll \frac{1 - t\frac{k}{z}}{1 -\frac{k}{z}} \rr
\\ \
[\Psis(k),H_{-}(z,t)] 
&=& 
\Psis(k) \sum_{n \in \mathbb{N}} \frac{1-t^n}{n} (zk)^n
=
\Psis(k) \log \ll \frac{1 - tzk}{1 - zk} \rr
\end{eqnarray}

These equations can be cast into the form

\begin{eqnarray}
e^{-H_{+}(z,t)} \Psi(k) e^{H_{+}(z,t)}
&=&
\Psi(k) \ll \frac{1 - t\frac{k}{z}}{1 -\frac{k}{z}} \rr
\\
e^{-H_{-}(z,t)} \Psis(k) e^{H_{-}(z,t)}
&=&
\Psis(k) \ll \frac{1 - tzk}{1 - zk} \rr
\end{eqnarray}

By equating coefficients of powers in $k$, we have the following $t$-fermion
evolution equations 

\begin{eqnarray}
\Gamma_{+}^{-1}(z,t) \psi_m \Gamma_{+}(z,t) &=& 
\psi_m + (1-t) \sum_{j=1}^{\infty} \psi_{(m-j)} z^{-j}
\label{h-v1}
\\
\Gamma_{-}(z,t) \psis_m \Gamma_{-}^{-1}(z,t) &=& 
\psis_m + (1-t) \sum_{j=1}^{\infty} \psis_{(m-j)} z^j
\label{h-v2}
\end{eqnarray}

\noindent where we have used the fact that $e^{-H_{+}(z,t)} = \Gamma_{+}^{-1}(z,t)$
and $e^{-H_{-}(z,t)} = \Gamma_{-}(z,t)$.

\subsection{\lq Fake\rq\  and \lq genuine\rq\  vacuum states}

Following \cite{mjd}, we introduce two types of (left and right) 
vacuum states. The {\it \lq fake\rq} vacuum states are represented 
by $\ivac$ and $\fvac$ and satisfy an inner product normalised to
$\ivac\Omega\rangle = 1$. The {\it \lq genuine\rq} vacuum states, 
$\idirac$ and $\fdirac$, are generated by infinite strings of 
$t$-fermion operators acting on $\ivac$ and $\fvac$ as follows 

\begin{eqnarray}
\idirac = \ivac \ldots \psi_2 \psi_1 \psi_0 
&=& 
\ivac \lprod_{j=0\phantom{|}}^{\phantom{|}\infty\phantom{?}} \psi_j
\label{idirac} 
\\
\fdirac = \psis_0 \psis_1 \psis_2 \ldots \fvac 
&=& 
\rprod_{j=0\phantom{|}}^{\infty\phantom{|}} \psis_j \fvac
\label{fdirac}
\end{eqnarray}

By convention, the energy of $\idirac$ and $\fdirac$ is set 
to zero \cite{mjd}. 

\subsection{Finite energy states}

Finite energy states are of the form
$\ivac \ldots \psi_{m_2} \psi_{m_1} \psi_{m_0}$ 
and the form
$\psis_{m_0}\psis_{m_1}\psis_{m_2} \ldots \fvac$, 
where $m_n = n$ for all $n \geq N$, for some sufficiently large $N$.
They differ from the vacuum states by a relabelling 
of a finite number of $t$-fermions with new indices. 

\subsection{The annihilation $t$-fermion operators}

We fix
\begin{eqnarray}
\ivac \ldots \psi_{(l+2)} \psi_{(l+1)} \psi_{l} \psis_m &=&
\langle l| \psis_m \ = \ 0
\label{annihilation1}
\\
\psi_m \psis_l \psis_{(l+1)} \psis_{(l+2)} \ldots \fvac &=&
\psi_m |l\rangle \ = \ 0
\label{annihilation2}
\end{eqnarray}

\noindent for all $m<l$. There are other annihilation operators
as well. In particular, 

\begin{equation}
\psis_m |l\rangle = 0, \quad 
\langle l| \psi_m = 0
\label{annihilation3}
\end{equation}

\noindent for all $m \geq l$.  
It is sufficient for our purposes to impose 
(\ref{annihilation1}--\ref{annihilation2}) as extra conditions 
and note that they are consistent with (\ref{a-c1}--\ref{a-c3}). 
(\ref{annihilation3}) follows from (\ref{a-c1}--\ref{a-c2}).

\section{State vectors and Young diagrams}
\label{P}

\subsection{Young diagrams label state vectors}

We use Young diagrams to label (left and right) state vectors in 
the (left and right) Fock spaces. The empty diagram 
labels $\idirac$ and also $\fdirac$. Non-empty diagrams label 
finite energy states.  If $\mu = \{\mu_1, \ldots, \mu_l\}$ is 
a diagram with $l$ non-zero parts, it labels the left and right 
state vectors

\begin{eqnarray}
\langle\mu| &=& 
\ivac 
\ldots 
\psi_{(l+1)} \psi_{l} \psi_{m_l} 
\ldots 
\psi_{m_1}
\\
|\mu\rangle &=&
\psis_{m_1} \ldots \psis_{m_l} \psis_{l} \psis_{(l+1)} 
\ldots 
\fvac
\label{partition}
\end{eqnarray}

\noindent where the integers $\{m_1,\ldots,m_l\}$ are given by
$m_j=j-1-\mu_j$, for $1 \leq j \leq l$. Notice that for a Young
diagram
with $l$ parts, the $l$ $t$-fermions that are farthest from the
corresponding fake vacuum state are relabelled.

\subsection{Lemma (Inner products)}

Left and right state vectors are orthogonal in the sense that

\begin{equation}
\label{innerproduct}
\langle \mu | \nu \rangle = b_{\mu}(t) \delta_{\mu,\nu}
\end{equation}

\noindent where $b_{\mu}(t)$ is defined as follows. If $\mu$ 
has $p_j(\mu)$ parts of length $j$, where $j \geq 1$, then

\begin{equation}
b_{\mu}(t)
=
\prod_{j=1}^{\infty}
\ll
\prod_{k=1}^{p_j(\mu)}
(1-t^k)
\rr
\end{equation}

\subsubsection{Proof.}

Consider the scalar product

\begin{equation}
\langle \mu | \nu \rangle
=
\langle \Omega| 
\ldots \psi_l \psi_{\tilde{m}_l} \ldots \psi_{\tilde{m}_1} \psis_{m_1}
\ldots \psis_{m_l} \psis_l \ldots 
|\Omega \rangle
\end{equation}

{}From (\ref{a-c4}) and (\ref{annihilation1}--\ref{annihilation2}), 
$\langle \mu | \nu \rangle = 0$ unless $\tilde{m}_1=m_1$, hence 
we set $\tilde{m}_1=m_1$ without loss of generality. Furthermore, 
let $m_1, \ldots, m_s$ be nearest neighbours for $1 \leq s \leq l$. 
That is, assume that $m_{j+1} = m_j + 1$ for $1 \leq j \leq (s-1)$, 
but fix $m_{s+1} > m_s+1$. Commuting 
$\psi_{\tilde{m}_1} \psis_{m_1}$ using (\ref{a-c4}), we obtain 

\begin{eqnarray}
\langle \mu | \nu \rangle
&=&
(1-t)
\langle \Omega| 
\ldots \psi_l \psi_{\tilde{m}_l} \ldots \psi_{\tilde{m}_2} \psis_{m_2}
\ldots \psis_{m_l} \psis_l \ldots |\Omega \rangle
\\
&+&
t
\langle \Omega| \ldots \psi_l \psi_{\tilde{m}_l} \ldots \psi_{\tilde{m}_3} \psi_{\tilde{m}_2}
\psis_{m_2}\psi_{m_2} \psis_{m_2} \psis_{m_3} \ldots \psis_{m_l} \psis_l
\ldots |\Omega \rangle
\nonumber
\\
&\vdots&
\nonumber
\\
&=&
(1-t^s)
\langle \Omega| \ldots \psi_l \psi_{\tilde{m}_l} \ldots \psi_{\tilde{m}_2} \psis_{m_2}
\ldots \psis_{m_l} \psis_l \ldots |\Omega \rangle
\nonumber
\end{eqnarray}

Repeating this procedure, we find that $\langle \mu | \nu \rangle = 0$
unless $\tilde{m}_j=m_j$ for all $1 \leq j \leq s$. Continuing to
commute the central pair of $t$-fermions $\psi_{\tilde{m}_j} \psis_{m_j}$ 
we find

\begin{equation}
\langle \mu | \nu \rangle
=
\prod_{k=1}^{s} (1-t^k)
\times
\langle \Omega| \ldots \psi_l \psi_{\tilde{m}_l} \ldots \psi_{\tilde{m}_{(s+1)}}
\psis_{m_{(s+1)}} \ldots \psis_{m_l} \psis_l \ldots |\Omega \rangle
\end{equation}

Thus we have acquired a factor of $\prod_{k=1}^{s} (1-t^k)$ for 
a set of $s$ parts of the same length in the Young diagram $\mu$.
The required result follows inductively.

\subsection{Normalization of the genuine vacuum states}

{}From (\ref{innerproduct}), we obtain the normalization 
$\langle 0|0\rangle = 1$, which is non-trivial to compute 
directly from $\langle \Omega|\Omega\rangle = 1$, given the 
nature of the $t$-anti-commutators (\ref{a-c1}--\ref{a-c3}).

\subsection{Interlacing Young diagrams}

Let $\lambda = \{\lambda_1,\ldots,\lambda_{(l+1)} \}$ 
be a diagram consisting of at least $l$ non-zero parts, that 
is, we allow $\lambda_{(l+1)}=0$. We say that $\lambda$ interlaces 
with $\mu$, where 
$\mu = \{\mu_1, \ldots, \mu_l\}$, 
and write 
$\lambda \succ \mu$ if $\lambda_j \geq \mu_j \geq \lambda_{(j+1)}$, 
for all $1 \leq j \leq l$. 

\subsection{Interlacing state vectors}

Since Young diagrams label state vectors, we can define interlacing 
state vectors as follows. In terms of right state vectors, if 

\begin{eqnarray}
|\lambda\rangle &=& 
\psis_{n_1}\ldots \psis_{n_l} \psis_{n_{(l+1)}} 
\psis_{(l+1)}\psis_{(l+2)} \ldots\fvac
\label{lambda} 
\\
|\mu\rangle &=&
\psis_{m_1}\ldots \psis_{m_l} 
\psis_{l} \psis_{(l+1)}\psis_{(l+2)} \ldots\fvac
\label{mu}
\end{eqnarray}

\noindent and $n_j \leq m_j \leq n_{(j+1)}-1$, for all 
$1 \leq j \leq l$, we say 
$|\lambda\rangle \succ |\mu\rangle$. 
An obvious similar definition holds for left state vectors.

\section{Action of $t$-vertex operators on state vectors}
\label{I}

\subsection{Skew Hall-Littlewood functions}
Following \cite{mac}, $P_{\lambda\slash\mu}(z,t)$, the skew 
Hall-Littlewood function of a single variable $z$, indexed by 
the skew Young diagram $\lambda\slash\mu$, is 
\begin{equation}
P_{\lambda\slash\mu}(z,t) 
=
\left\{
\begin{array}{cc}
\Phi_{\lambda\slash\mu}(t) z^{|\lambda|-|\mu|} & \lambda \succ \mu
\\
\\
0 & \lambda \not\succ \mu
\end{array} 
\right.
\end{equation}
\noindent where $\Phi_{\lambda\slash\mu}(t)$ is a polynomial in $t$ 
which we now describe. Let $\mu$ have $p_j(\mu)$ parts of 
length $j$ and $\lambda$ have $p_j(\lambda)$ parts of
length $j$. Define $J_{\lambda\slash\mu}$ to be the set of 
integers $j$ such that $p_j(\lambda)-p_j(\mu) = -1$. Then 
$\Phi_{\lambda\slash\mu}(t)$ is 
\begin{equation}
\Phi_{\lambda\slash\mu}(t)
=
\prod_{j \in J_{\lambda\slash\mu}} (1-t^{p_j(\mu)})
\end{equation}
That is, when the number of parts $p_j(\mu)$ of length $j$ 
decreases by 1 in the transition from $\mu$ to $\lambda$, 
a factor of $(1-t^{p_j(\mu)})$ is acquired. 

\subsection{Lemma (Action of $t$-vertex operators)}

We prove the following results

\begin{eqnarray}
\quad
\Gamma_{-}(z,t) |\mu\rangle
=
\sum_{\lambda \succ \mu} 
P_{\lambda\slash\mu}(z,t) 
|\lambda\rangle
=
\sum_{\lambda \succ \mu}
\ll
\prod_{j \in J_{\lambda\slash\mu}} (1-t^{p_j(\mu)})
z^{|\lambda|-|\mu|}
\rr
|\lambda\rangle
\label{lemma1}
\\
\quad
\langle\mu| 
\Gamma_{+}(z,t)
= 
\sum_{\lambda \succ \mu} 
P_{\lambda\slash\mu}(z^{-1},t) 
\langle\lambda|
=
\sum_{\lambda \succ \mu} 
\ll
\prod_{j \in J_{\lambda\slash\mu}} (1-t^{p_j(\mu)})
z^{|\mu|-|\lambda|}
\rr
\langle\lambda|
\label{lemma2}
\end{eqnarray}
\phantom{.}

\noindent where the sums are over all Young diagrams $\lambda$ 
which interlace with $\mu$. This lemma is the first of two
results in this work. We prove (\ref{lemma1}) in detail. 
The proof of (\ref{lemma2}) is analogous. We require the 
following identities.

\subsection{Identities}

Firstly, from (\ref{a-c5}), we can derive

\begin{eqnarray}
\label{id1}
&\phantom{=}&
\ll \psis_m + (1-t) \sum_{j=1}^{\infty} \psis_{(m-j)}z^j \rr
\ll \sum_{j=0}^{\infty} \psis_{(m-j)} z^j \rr
\\
&=&
\sum_{n=0}^{\infty} \ll \psis_{m}\psis_{(m-n)} + (1-t) \sum_{j=1}^{n}
\psis_{(m-j)}\psis_{(m+j-n)} \rr
z^n
\nonumber
\\
&=&
t
 \sum_{n=0}^{\infty} \psis_{(m-1-n)}
\psis_{(m+1)}
z^n
\nonumber
\end{eqnarray}

Next, we use (\ref{id1}) to derive two more identities.

\begin{eqnarray}
\label{trunc-1}
&&
\ll
\psis_m + (1-t)\sum_{j=1}^{\infty}\psis_{(m-j)} z^j
\rr
\ll
\psis_{(m+1)} + (1-t^p)\sum_{j=1}^{\infty} \psis_{(m+1-j)} z^j
\rr
\\
&=&
\ll
\psis_m + (1-t)\sum_{j=1}^{\infty}\psis_{(m-j)} z^j
\rr
\psis_{(m+1)}
\nonumber
\\
&+&
t (1-t^p)
\ll
\sum_{j=1}^{\infty} \psis_{(m-j)}z^j
\rr
\psis_{(m+1)}
\nonumber
\\
&=&
\ll
\psis_m + (1-t^{p+1})\sum_{j=1}^{\infty}\psis_{(m-j)} z^j
\rr
\psis_{(m+1)}
\nonumber
\end{eqnarray}

\begin{eqnarray}
\label{trunc-2}
& &
\ll
\psis_m + (1-t)\sum_{j=1}^{\infty}\psis_{(m-j)} z^j
\rr
\ll
\psis_{(m+n)} + (1-t^p)\sum_{j=1}^{\infty} \psis_{(m+n-j)} z^j
\rr
\\
\nonumber
&=&
\ll
\psis_m + (1-t)\sum_{j=1}^{\infty}\psis_{(m-j)} z^j
\rr
\ll
\psis_{(m+n)} + (1-t^p)\sum_{j=1}^{n-1}\psis_{(m+n-j)}z^j
\rr
\\
\nonumber
&+&
t(1-t^p)
z^{n-1}
\ll
\sum_{j=1}^{\infty} \psis_{(m-j)} z^j
\rr
\psis_{(m+1)}
\\
\nonumber
&=&
\ll
\psis_m + (1-t)\sum_{j=1}^{\infty}\psis_{(m-j)} z^j
\rr
\ll
\psis_{(m+n)} + (1-t^p)\sum_{j=1}^{n-2}\psis_{(m+n-j)}z^j
\rr
\\
\nonumber
&+&
(1-t^p)
z^{n-1}
\ll
\psis_m + \sum_{j=1}^{\infty}\psis_{(m-j)} z^j
\rr
\psis_{(m+1)}
\end{eqnarray}

\noindent where $p \geq 1$ may be infinite, $n \geq 2$, and 
(\ref{id1}) was used between the first and second lines in 
(\ref{trunc-1}) and (\ref{trunc-2}).

\subsection{Notation}

We introduce the notation

\begin{equation}
\ll
\psis_m + (1-x) \sum_{j=1}^{\infty} \psis_{(m-j)} z^j
\rr
= \Psis_m(z,x)
\end{equation}

\noindent which allows us to write (\ref{trunc-1}--\ref{trunc-2}) 
in the succinct form

\begin{eqnarray}
\Psis_m(z,t) \Psis_{(m+1)}(z,t^p) = \Psis_m(z,t^{p+1})\psis_{(m+1)}
\label{trunc-3}
\end{eqnarray}

\begin{eqnarray}
\quad\quad
\Psis_m(z,t) \Psis_{(m+n)}(z,t^p)
&=&
\Psis_m(z,t)
\ll
\psis_{(m+n)} + (1-t^p)\sum_{j=1}^{n-2}\psis_{(m+n-j)}z^j
\rr
\label{trunc-4}
\\
&+&
(1-t^p) z^{n-1}\Psis_m(z,t^{\infty}) \psis_{(m+1)}
\nonumber
\end{eqnarray}

\noindent where we are defining $t^{\infty} = 0$, in view of the fact that
$|t| < 1$, as described in subsection {\bf 1.1}.

\subsection{Proof}

We are now in a position to outline the proof of (\ref{lemma1}).
The proof consists of three parts. 
{\bf 1.} We show that the action of $\Gamma_{-}$ on a right 
state vector labelled by a diagram $\mu$ generates a weighted 
sum over all right state vectors labelled by Young diagrams 
$\lambda \succ \mu$. 
{\bf 2.} We show that $\Phi_{\lambda\slash\mu}(t)$, the first 
factor in $P_{\lambda\slash\mu}(z, t)$, is recovered. 
{\bf 3.} We show that $z^{|\lambda|-|\mu|}$, 
the second factor in $P_{\lambda\slash\mu}(z, t)$, 
is recovered.

\subsubsection{Part 1: Generating all interlacing Young diagrams}
 
We start from expression $(\ref{partition})$ for $|\mu\rangle$. 
We act on this right state vector with $\Gamma_{-}(z,t)$ and 
make repeated use of the relation (\ref{h-v2}), to obtain 

\begin{eqnarray}
\label{h-v4}
& & \Gamma_{-}(z,t) |\mu\rangle =
\Gamma_{-}(z,t)
\psis_{m_1} \ldots \psis_{m_l} \psis_{l} \psis_{(l+1)}
\ldots
\fvac
\\
\nonumber
\\
\nonumber
&=&
\ll
\Gamma_{-}\psis_{m_1}\Gamma_{-}^{-1} 
\rr
\ldots
\ll
\Gamma_{-}\psis_{m_l}\Gamma_{-}^{-1} 
\rr
\ll
\Gamma_{-}\psis_l\Gamma_{-}^{-1} 
\rr
\ll
\Gamma_{-}\psis_{(l+1)}\Gamma_{-}^{-1} 
\rr
\ldots
\fvac
\\
\nonumber
&=&
\ll
\psis_{m_1}+(1-t)\sum_{j=1}^{\infty}\psis_{(m_1-j)}z^j
\rr
\ldots
\ll
\psis_{m_l}+(1-t)\sum_{j=1}^{\infty}\psis_{(m_l-j)}z^j
\rr
\\
\nonumber
&\times&
\ll
\psis_{l}+(1-t)\sum_{j=1}^{\infty}\psis_{(l-j)}z^j
\rr
\ll
\psis_{(l+1)}+(1-t)\sum_{j=1}^{\infty}\psis_{(l+1-j)}z^j
\rr
\ldots
\fvac
\\
\nonumber
&=& 
\Psis_{m_1}(z,t)
\ldots
\Psis_{m_l}(z,t)
\Psis_{l}(z,t)
\Psis_{(l+1)}(z,t)
\ldots
\fvac
\end{eqnarray}

Using (\ref{trunc-3}) to truncate all but the first $(l+1)$ sums 
appearing in (\ref{h-v4}), we obtain

\begin{eqnarray}
\Gamma_{-}(z,t) |\mu\rangle
=
\overbrace{\Psis_{m_1}(z,t)}^{n_1}
\ldots
\overbrace{\Psis_{m_l}(z,t)}^{n_l}
\overbrace{\Psis_{l}(z,t^{\infty})}^{n_{(l+1)}}
\psis_{(l+1)}
\psis_{(l+2)}
\ldots
\fvac
\label{something}
\end{eqnarray}

The proof requires that the first $(l+1)$ sums truncate to give 
exactly all Young diagrams of the form $(\ref{lambda})$. To see that 
this is the case, we use (\ref{trunc-3}--\ref{trunc-4}) 
iteratively. Firstly, one computes the product of sums marked 
$n_l$ and $n_{(l+1)}$, using (\ref{trunc-4}) with $p=\infty$,
to obtain 

\begin{eqnarray}
&&\Gamma_{-}(z,t) |\mu\rangle
\ \
=
\label{trunc-6}
\\
&&
\overbrace{\Psis_{m_1}(z,t)}^{n_1}
\ldots
\overbrace{\Psis_{m_{(l-1)}}(z,t)}^{n_{(l-1)}}
\overbrace{\Psis_{m_l}(z,t)}^{n_l}
\overbrace{
\ll
\psis_l + \sum_{j=1}^{l-2-m_l} \psis_{(l-j)} z^j 
\rr
}^{n_{(l+1)}}
\psis_{(l+1)}
\ldots
\fvac
\nonumber
\\
&+&
z^{l-1-m_l}
\overbrace{\Psis_{m_1}(z,t)}^{n_1}
\ldots
\overbrace{\Psis_{m_{(l-1)}}(z,t)}^{n_{(l-1)}}
\overbrace{\Psis_{m_l}(z,t^{\infty})}^{n_l}
\overbrace{\psis_{(m_l+1)}}^{n_{(l+1)}}
\psis_{(l+1)}
\ldots
\fvac
\nonumber
\end{eqnarray} 

Notice that we have truncated the $n_{(l+1)}$ sum, so that only 
$t$-fermions $\psis_{n_{(l+1)}}$ with indices 
$m_l+1 \leq n_{(l+1)} \leq l$ remain. These are precisely the 
permissible values for $n_{(l+1)}$ in $(\ref{lambda})$ so that 
$\lambda \succ \mu$.

Now if $m_l-m_{(l-1)} =1$ ($m_l-m_{(l-1)} \geq 2$), take the product 
of the sums marked $n_{(l-1)}$ and $n_l$ in (\ref{trunc-6}), using 
(\ref{trunc-3}) (using (\ref{trunc-4})) with $p=1$ for the first term 
and $p=\infty$ for the second. This truncates the $n_l$ terms, so that 
only $t$-fermions $\psis_{n_l}$ with indices 
$m_{(l-1)} + 1 \leq n_l \leq m_l$ remain. Again, these are precisely 
the permissible values for $n_l$ in (\ref{lambda}) so that 
$\lambda \succ \mu$. 

Continuing this way, the truncation ultimately extends to give 
$n_1 \leq m_1$ and $m_j + 1 \leq n_{(j+1)} \leq m_{(j+1)}$, for all 
$1 \leq j \leq l-1$. That is, we recover exactly all Young
diagrams of 
the form $(\ref{lambda})$ from (\ref{trunc-6}), proving that 
$\Gamma_{-}(z,t)$ generates exactly all $\lambda \succ \mu$.

\subsubsection{Part 2: Factor $\Phi_{\lambda\slash\mu}(t)$ of 
$P_{\lambda\slash\mu}(z,t)$}

For $1 \leq a \leq b \leq l$, let $\{m_{a},\ldots,m_{b}\}$ be 
a contiguous subset of the points $\{m_1,\ldots,m_l\}$, which 
describe the Young diagram $\mu$ in (\ref{partition}). In other 
words, the points $\{m_{a},\ldots,m_{b}\}$ are nearest neighbours. 
These points will give rise to $(b-a+1)$ parts in $\mu$ of a certain 
length $h$. Under the action of $\Gamma_{-}$ the points 
$\{ m_a,\ldots, m_b,m_{(b+1)} \}$ are shifted to 
$\{ n_a, \ldots n_b, n_{(b+1)}\}$, where we define $m_{(l+1)} = l$ 
if $b=l$. Because this shifting must produce interlacing Young
diagrams 
$\lambda$, we can have $m_{(a-1)}+1 \leq n_a \leq m_a$, where 
$m_0 = -\infty$ when $a=1$, and $m_b+1 \leq n_{(b+1)} \leq m_{(b+1)}$, 
but all other indices are stationary.

There are three cases to consider. {\bf 1.} When $n_a = m_a$ and 
$n_{(b+1)} \not= m_b+1$, there will still be exactly $(b-a+1)$ parts 
of length $h$ in the new Young diagram $\lambda$, so no weight is 
expected. 
{\bf 2.} When $n_a < m_a$ and $n_{(b+1)} \not= m_b + 1$, there will 
be exactly $(b-a)$ parts of length $h$ in $\lambda$, meaning 
$p_h(\lambda) - p_h(\mu)=-1$. Consequently, we expect a weight 
of $(1-t^{b-a+1})$ to be acquired. 
{\bf 3.} When $n_{(b+1)} = m_b+1$, there will be at least $(b-a+1)$ 
parts of length $h$ in $\lambda$, so no weight is expected. 

Let us check that these weights are in fact recovered, 
by analysing a general term which arises after repeated truncation 
of (\ref{something}). In the following we abbreviate 
$\Psis_m(z,t) = \Psis_m$, for the sake of visual clarity. Using 
(\ref{trunc-4}), we have

\begin{eqnarray}
\label{something2}
\\ 
&\ldots&\Psis_{m_{(a-1)}}
\Psis_{m_a}
\ldots
\Psis_{m_b}
\Psis_{m_{(b+1)}}(z,t^p)
\ldots
\fvac =
\nonumber
\\
\nonumber
\\
&\ldots&\Psis_{m_{(a-1)}}
\Psis_{m_a}
\ldots
\Psis_{m_b}
\ll
\psis_{m_{(b+1)}}
+
(1-t^p)
\sum_{j=1}^{m_{(b+1)}-m_b-2}
\psis_{(m_{(b+1)}-j)}z^j
\rr
\ldots
\fvac
\nonumber
\\
\nonumber
\\
&+&
\ldots
(1-t^p)
z^{m_{(b+1)}-m_b-1}
\Psis_{m_{(a-1)}}
\Psis_{m_a}
\ldots
\Psis_{m_b}(z,t^{\infty})
\psis_{(m_b+1)}
\ldots
\fvac
\nonumber
\end{eqnarray}

Using (\ref{trunc-3}) repeatedly in both the first and second terms 
in (\ref{something2}), leads to 

\begin{eqnarray}
&&
\ldots 
\Psis_{m_{(a-1)}}
\Psis_{m_a}
\ldots
\Psis_{m_b}
\Psis_{m_{(b+1)}}(z,t^p)
\ldots
\fvac
\\
\nonumber
\\
&=&
\ldots \Psis_{m_{(a-1)}}
\Psis_{m_a}(z,t^{b-a+1})
\psis_{m_{(a+1)}}
\ldots
\psis_{m_b}
\nonumber
\\
&\times&
\ll
\psis_{m_{(b+1)}}
+
(1-t^p)
\sum_{j=1}^{m_{(b+1)}-m_b-2}
\psis_{(m_{(b+1)}-j)}z^j
\rr
\ldots
\fvac
\nonumber
\\
\nonumber
\\
&+&
\ldots
(1-t^p)
z^{m_{(b+1)}-m_b-1}
\Psis_{m_{(a-1)}}
\Psis_{m_a}(z,t^{\infty})
\psis_{m_{(a+1)}}
\ldots
\psis_{m_b}
\psis_{(m_b+1)}
\ldots
\fvac
\nonumber
\end{eqnarray}

The above calculation has split the original object into two terms. 
The first term features Young diagrams for which $n_{(b+1)} \not= m_b+1$. 
Hence, it has some Young diagrams $\lambda$ for which 
$p_h(\lambda) < p_h(\mu)$. Taking the product 
$\Psis_{m_{(a-1)}} \Psis_{m_a}(z,t^{b-a+1})$ using $(\ref{trunc-4})$, 
we find that we get no weight for $n_a = m_a$, whilst we acquire 
the factor $(1-t^{b-a+1})$ for $n_a < m_a$. This is exactly as required. 

The second term features Young diagrams for which $n_{(b+1)} = m_b+1$. Hence 
it only has Young diagrams $\lambda$ for which $p_h(\lambda) \geq p_h(\mu)$. 
Taking the product $\Psis_{m_{(a-1)}} \Psis_{m_a}(z,t^{\infty})$ using 
(\ref{trunc-4}), we find that we get no weight for any value of $n_a$. 
Again, this is the expected result.

Note that the factors of $(1-t^p)$ are weights due to the motion of 
$m_{(b+1)}$, which belongs to a different set of contiguous points.

\subsubsection{Part 3: Factor $z^{|\lambda|-|\mu|}$ of 
$P_{\lambda\slash\mu}(z,t)$}

If there are $\delta$ more boxes in the Young diagram of $\lambda$ than 
the Young diagram of $\mu$, this is because the indices of the 
$t$-fermions 
have been shifted downward by a net $\delta$ units. However, from 
(\ref{h-v2}) we see that a net downward shift by $\delta$ units will 
acquire the correct weight of $z^\delta$.

\subsubsection{Action of $t$-vertex operators on the left state 
vectors}

Using precisely the same steps as above, one derives the result 
in (\ref{lemma2}).

\section{Hall-Littlewood weighted plane partitions}
\label{PP}

\subsection{Plane partitions as 3-dimensional objects}

One can think of a plane partition $\pi$ as a set of unit cubic 
boxes stacked in the north-west corner of a 3-dimensional room 
such that the heights of columns of boxes are weakly decreasing 
as one moves horizontally farther from the corner \cite{mac}. A 
3-dimensional view of a plane partition is shown in Figure 
{\bf \ref{3D}}.

%
\begin{center}
\begin{minipage}{3.5in}
\setlength{\unitlength}{0.001cm}
\renewcommand{\dashlinestretch}{30}
\begin{picture}(4800, 4800)(0, 0)
\thicklines
\path(3900,1500)(4500,1500)
\path(1800,4200)(2400,4200)
\path(1800,4200)(1500,3900)
\path(2400,4200)(2100,3900)
\path(1500,3900)(2100,3900)
\path(1500,3900)(1500,3300)
\path(2100,3900)(2100,3300)
\path(1500,3300)(2100,3312)
\path(2400,4200)(2400,3600)
\path(2400,3600)(2100,3300)
\path(1500,3300)(1200,3000)
\path(2100,3300)(1800,3000)
\path(1200,3000)(1800,3000)
\path(1200,3000)(1200,2400)
\path(1800,3000)(1800,2400)
\path(1200,2400)(1800,2400)
\path(1200,2400)(0900,2100)
\path(1800,2400)(1500,2100)
\path(0900,2100)(1500,2100)
\path(0900,2100)(0900,1500)
\path(1500,2100)(1500,1500)
\path(0900,1500)(1500,1500)
\path(0900,1500)(0600,1200)
\path(1500,1500)(1200,1200)
\path(0600,1200)(1200,1200)
\path(0600,1200)(0600,0600)
\path(1200,1200)(1200,0600)
\path(0600,0600)(1200,0600)
\path(0600,0600)(0000,0000)
\path(2100,3300)(2100,2700)
\path(2100,2700)(1800,2400)
\path(1800,2400)(1800,1800)
\path(1800,1800)(1500,1500)
\path(1500,1500)(1500,0900)
\path(1500,0900)(1200,0600)
\path(2400,3600)(2400,3000)
\path(2400,3000)(2100,2700)
\path(2400,3000)(3000,3000)
\path(2100,2700)(2700,2700)
\path(3000,3000)(2700,2700)
\path(1800,2400)(2400,2400)
\path(2700,2700)(2400,2400)
\path(2400,2400)(2400,1800)
\path(1800,1800)(2400,1800)
\path(2400,1800)(2100,1500)
\path(1500,1500)(2100,1500)
\path(2100,1500)(2100,0900)
\path(1500,0900)(2100,0900)
\path(3000,3000)(3000,2400)
\path(2700,2700)(2700,2100)
\path(3000,2400)(2400,1800)
\path(3000,2400)(4500,2400)
\path(3600,2400)(2700,1500)
\path(2700,2100)(3300,2100)
\path(2400,1800)(3000,1812)
\path(2100,1500)(2700,1500)
\path(2700,1500)(2700,0900)
\path(2100,0900)(2700,0900)
\path(4200,2400)(3900,2100)
\path(4500,2400)(4800,2400)
\path(4800,2400)(4500,2100)
\path(3300,2100)(4500,2100)
\path(3000,1800)(3000,1200)
\path(3900,2100)(3900,1500)
\path(4500,2100)(4500,1500)
\path(4800,2400)(4800,1800)
\path(4800,1800)(4500,1500)
\path(4800,1800)(5400,1800)
\path(3900,2100)(3600,1800)
\path(3000,1800)(3600,1800)
\path(3600,1800)(3600,1200)
\path(3000,1200)(3600,1200)
\path(3900,1500)(3600,1200)
\path(3000,1200)(2700,0900)
\path(1800,4800)(1800,4200)
\end{picture}
\begin{ca}
\label{3D}
A 3-dimensional view of a plane partition. 
\end{ca}
\end{minipage}
\end{center}
\bigskip

\subsection{Plane partitions as 2-dimensional arrays}

One can also think of a plane partition $\pi$ as a 2-dimensional 
array (in the shape of a Young diagram) of non-negative integers 
$\pi_{i,j}$ on a 2-dimensional grid in the south-east quadrant 
of the plane. The 2-dimensional base of the plane partition $\pi$ 
of Figure {\bf \ref{3D}} is shown in Figure {\bf \ref{2D}}, 
together with its cell coordinates. 

The integers $\pi_{i,j}$, which correspond to the column 
heights in the 3-dimensional view of $\pi$, satisfy 

\begin{equation}
\pi_{i,j} \geq \pi_{i+1,j}, 
\quad 
\pi_{i,j} \geq \pi_{i,j+1},
\quad 
\lim_{i \rightarrow \infty} \pi_{i,j} = 
\lim_{j \rightarrow \infty} \pi_{i,j} = 0
\end{equation}

\noindent for all integers $i,j \geq 0$. 

\subsection{Slicing a plane partition diagonally}

As observed in \cite{or2}, when plane partitions are sliced 
diagonally, one obtains a sequence of interlacing Young diagrams. 
In other words, if for $k \geq 1$ we define the Young diagrams 

\begin{eqnarray}
\mu_{-k} 
& = & 
\{\pi_{k,0}, \pi_{(k+1),1}, \pi_{(k+2),2}, \ldots \}
\label{slices1}
\\
\mu_0 
&=& 
\{\pi_{0,0}, \pi_{1,1}, \pi_{2,2}, \ldots \}
\nonumber
\\
\mu_{k} 
&=& 
\{\pi_{0,k}, \pi_{1,(k+1)}, \pi_{2,(k+2)}, \ldots \} 
\nonumber
\end{eqnarray}

\noindent then every plane partition $\pi$ satisfies

\begin{equation}
\pi
\ = \
\{ 
\emptyset = 
\mu_{-m} \prec \cdots \prec 
\mu_{-1} \prec \mu_0 \succ 
\mu_1 \succ \cdots \succ 
\mu_n = \emptyset \}
\label{slicing}
\end{equation}

\noindent for some $m,n \geq 1$. A 2-dimensional view of the 
plane partition of Figure {\bf \ref{3D}}, together with its 
column heights and diagonal slices, is shown in Figure 
{\bf \ref{2D}}.

%
\begin{center}
\begin{minipage}{4.9in}
\setlength{\unitlength}{0.0012cm}
\renewcommand{\dashlinestretch}{30}
\begin{picture}(4800, 3600)(-3000, 0)
\thicklines
\path(0000,0600)(0600,0600)
\path(0000,1200)(1800,1200)
\path(0000,1800)(2400,1800)
\path(0000,2400)(3000,2400)
\path(0000,3000)(0000,0600)
\path(0000,3000)(3000,3000)
\path(0600,3000)(0600,0600)
\path(1200,3000)(1200,1200)
\path(1800,3000)(1800,1200)
\path(2400,3000)(2400,1800)
\path(3000,3000)(3000,2400)
\thinlines
\path(0000,1200)(0900,0300)
\path(0000,1800)(0900,0900)
\path(0000,2400)(1500,0900)
\path(0000,3000)(2100,0900)
\path(0600,3000)(2100,1500)
\path(1200,3000)(2700,1500)
\path(1800,3000)(2700,2100)
\path(2400,3000)(3300,2100)
\put(0300,0900){1}
\put(0300,1500){2}
\put(0300,2100){3}
\put(0300,2700){4}
\put(0900,1500){1}
\put(0900,2100){2}
\put(0900,2700){2}
\put(1500,1500){1}
\put(1500,2100){1}
\put(1500,2700){1}
\put(2100,2100){1}
\put(2100,2700){1}
\put(2700,2700){1}
\put(0900,0600){$\mu_{-2}$}
\put(0900,0000){$\mu_{-3}$}
\put(1500,0600){$\mu_{-1}$}
\put(2100,1200){$\mu_{ 1}$}
\put(2100,0600){$\mu_{ 0}$}
\put(2700,1800){$\mu_{ 3}$}
\put(2700,1200){$\mu_{ 2}$}
\put(3300,1800){$\mu_{ 4}$}
\put(0050,0700){{\tiny 30}}
\put(0050,1300){{\tiny 20}}
\put(0050,1900){{\tiny 10}}
\put(0050,2500){{\tiny 00}}
\put(0650,1300){{\tiny 21}}
\put(0650,1900){{\tiny 11}}
\put(0650,2500){{\tiny 01}}
\put(1250,1300){{\tiny 22}}
\put(1250,1900){{\tiny 12}}
\put(1250,2500){{\tiny 02}}
\put(1850,1900){{\tiny 13}}
\put(1850,2500){{\tiny 03}}
\put(2450,2500){{\tiny 04}}
\end{picture}
\begin{ca}
\label{2D}
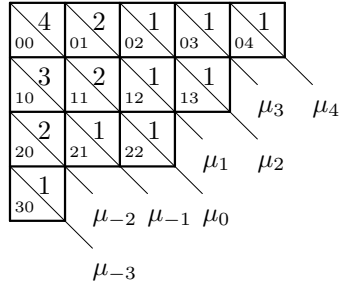
A 2-dimensional view of the plane partition in Figure {\bf \ref{3D}} 
and its diagonal slices. The upper (large sized) integers in the 
cells are the heights of the corresponding columns. The lower (small 
sized) integers are the $(i, j)$ cell coordinates.
\end{ca}
\end{minipage}
\end{center}
\bigskip

%
\begin{center}
\begin{minipage}{4.9in}
\setlength{\unitlength}{0.001cm}
\renewcommand{\dashlinestretch}{30}
\begin{picture}(4800, 5000)(-4000, 0)
\thicklines
%
\path(0000,0000)(0600,0000)
\path(0000,0600)(1800,0600)
\path(0000,1200)(3000,1200)
\path(0000,1800)(3600,1800)
\path(0000,2400)(3600,2400)
\path(0000,3000)(3600,3000)
\path(0000,3600)(4200,3600)
\path(0000,4200)(4200,4200)
%
%
\path(4200,4200)(4200,3600)
\path(3600,4200)(3600,1800)
\path(3000,4200)(3000,1200)
\path(2400,4200)(2400,1200)
\path(1800,4200)(1800,0600)
\path(1200,4200)(1200,0600)
\path(0600,4200)(0600,0000)
\path(0000,4200)(0000,0000)
\put(0150,3850){$5_3$}
\put(0150,3250){$5_2$}
\put(0150,2650){$5_1$}
\put(0150,2050){$4_2$}
\put(0150,1450){$4_1$}
\put(0150,0850){$2_1$}
\put(0150,0250){$1_1$}
\put(0750,3850){$5_2$}
\put(0750,3250){$5_2$}
\put(0750,2650){$5_1$}
\put(0750,2050){$4_1$}
\put(0750,1450){$4_1$}
\put(0750,0850){$2_1$}
\put(1350,3850){$5_1$}
\put(1350,3250){$5_1$}
\put(1350,2650){$5_1$}
\put(1350,2050){$4_1$}
\put(1350,1450){$3_1$}
\put(1350,0850){$2_1$}
\put(1950,3850){$5_1$}
\put(1950,3250){$4_1$}
\put(1950,2650){$3_2$}
\put(1950,2050){$3_1$}
\put(1950,1450){$2_1$}
\put(2550,3850){$4_1$}
\put(2550,3250){$3_1$}
\put(2550,2650){$3_1$}
\put(2550,2050){$3_1$}
\put(2550,1450){$1_1$}
\put(3150,3850){$2_1$}
\put(3150,3250){$2_1$}
\put(3150,2650){$2_1$}
\put(3150,2050){$1_1$}
\put(3750,3850){$1_1$}
\end{picture}
\begin{ca}
\label{plane-partition}
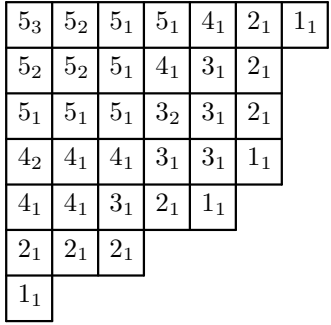
A 2-dimensional view of a plane partition. $H_m$ in each cell 
stand for height $H$ and level $m$. There are 13 level-1, 
3 level-2, and 1 level-3 paths. The associated weight is 
$A_{\pi}(t) = (1-t)^{13} (1-t^2)^{3} (1-t^3)$.
\end{ca}
\end{minipage}
\end{center}
\bigskip

\subsection{Levels and paths}

Consider the elements $\{\pi_{i,j}\}$ that form a plane partition 
array. The element $\pi_{i,j}$ with coordinates $(i,j)$ is 
contiguous with four surrounding elements at 
$\{(i-1,j),\ (i+1,j),\ (i,j-1),\ (i,j+1)\}$, possibly including 
trivial elements that have numerical value $0$ (trivial elements
of a plane partition that have height 0). 

Following \cite{v2}, we say that $\pi_{i, j}$ has level $l \geq 1$ 
if 

\begin{equation}
\pi_{i,j} 
= 
\ldots 
= 
\pi_{(i+l-1),(j+l-1)} > \pi_{(i+l),(j+l)}
\end{equation}

With these definitions, a path of level-$l$ is a set of contiguous 
level-$l$ elements that have the same numerical value. They correspond 
to equal-height columns in the 3-dimensional view of $\pi$. 

We say that the plane partition $\pi$ has $p_j(\pi)$ level-$j$ 
paths and assign it the weight $A_{\pi}(t)$, given by

\begin{equation}
\label{a-weight}
A_{\pi}(t) = \prod_{j=1}^{\infty} \ll 1-t^j\rr^{p_j(\pi)}
\end{equation}

An example of a larger plane partition is shown in Figure 
{\bf \ref{plane-partition}}. In this example, there are 3 
paths of column height 5. They consist of a level-1, 
a level-2 and a level-3 path. Equivalently, this height
5 horizontal plane (a set of contiguous paths of the same 
column height) has a maximal width of 3 paths, when measured 
diagonally. 

It is straightforward to show that

\begin{equation}
A_{\pi}(t) = 
b_{\mu_0}(t) 
\prod_{j=1}^{m}\Phi_{\mu_{(-j+1)}\slash \mu_{(-j)}}(t)
\prod_{k=1}^{n}\Phi_{\mu_{(k-1)} \slash \mu_{(k)}}(t)
\end{equation}

\noindent where the Young diagrams $\mu_j$ are the diagonal slices of 
$\pi$, as given by $(\ref{slicing})$.

\subsection{A scalar product generates weighted plane partitions} 
We study the infinite scalar product

\begin{eqnarray}
\label{scalar-product}
\mathcal{S}(t)
&=&
\idirac \ldots 
\Gamma_{+}(z^{-\frac{3}{2}},t) \Gamma_{+}(z^{-\frac{1}{2}},t)
\Gamma_{-}(z^{\frac{1}{2}},t) \Gamma_{-}(z^{\frac{3}{2}},t)
\ldots \fdirac
\label{scalar}
\\
\nonumber
\\
&=&
\idirac
\lprod_{j=1}^{\infty}
\Gamma_{+}(z^{\frac{-2j+1}{2}},t)
\rprod_{k=1}^{\infty}
\Gamma_{-}(z^{\frac{2k-1}{2}},t)
\fdirac
\nonumber
\end{eqnarray}

The scalar product $\mathcal{S}(t)$ provides a generating function for plane 
partitions. This is seen in the following way. We insert a complete 
set of states 
$\sum_{\mu_0} \frac{1}{b_{\mu_0}(t)}|\mu_0\rangle\langle\mu_0|$ 
between the central pair of vertex operators in $(\ref{scalar})$, 
giving

\begin{eqnarray}
\mathcal{S}(t)
=
\sum_{\mu_0}
\frac{1}{b_{\mu_0}(t)} 
\idirac
\lprod_{j=1}^{\infty}
\Gamma_{+}(z^{\frac{-2j+1}{2}},t)
|\mu_0\rangle\langle\mu_0|
\rprod_{k=1}^{\infty}
\Gamma_{-}(z^{\frac{2k-1}{2}},t)
\label{c}
\fdirac
\end{eqnarray}

{}From (\ref{lemma1}) for right state vectors, and (\ref{lemma2}) 
for left state vectors, it is clear that all plane partitions of the 
form (\ref{slicing}) receive a weight equal to

\begin{eqnarray}
&
b_{\mu_0}(t)
&
\prod_{j=1}^{m}\Phi_{\mu_{(-j+1)} \slash \mu_{(-j)}}(t)\ z^{\frac{2j-1}{2}(|\mu_{(-j+1)}|-|\mu_{(-j)}|)}
\\
&\times&
\prod_{k=1}^{n}\Phi_{\mu_{(k-1)} \slash \mu_{(k)}}(t)\ z^{\frac{2k-1}{2}(|\mu_{(k-1)}|-|\mu_{(k)}|)}
\nonumber
\\
&=&
\
A_{\pi}(t)z^{\sum_j |\mu_{j}|}
\ \
=
\ \
A_{\pi}(t)z^{|\pi|}
\nonumber
\end{eqnarray}

\noindent from (\ref{c}). In other words,

\begin{equation}
\label{gf-additive}
\mathcal{S}(t) = \sum_{\pi} A_{\pi}(t) z^{|\pi|}
\end{equation}

\subsection{Evaluating the scalar product directly}

On the other hand, it is possible to evaluate $\mathcal{S}(t)$ directly. 
Using the commutation relation $(\ref{h-v3})$ repeatedly, we find

\begin{eqnarray}
\nonumber
\mathcal{S}(t) 
&=&
\ll
\prod_{j=1}^{\infty}\frac{1-t z^j}{1-z^j} 
\rr
\idirac
\lprod_{j=1}^{\infty}
\Gamma_{+}(z^{\frac{-2j+1}{2}},t)
\rprod_{k=2}^{\infty}
\Gamma_{-}(z^{\frac{2k-1}{2}},t)
\fdirac
\\
\nonumber
&=&
\ll
\prod_{j=1}^{\infty} \frac{1-t z^j}{1-z^j} 
\rr
\ll
\prod_{k=2}^{\infty} \frac{1-t z^k}{1-z^k} 
\rr
\idirac
\lprod_{j=1}^{\infty}
\Gamma_{+}(z^{\frac{-2j+1}{2}},t)
\rprod_{k=3}^{\infty}
\Gamma_{-}(z^{\frac{2k-1}{2}},t)
\fdirac
\\
&=&
\prod_{j=1}^{\infty}
\ll
\frac{1-t z^j}{1-z^j}
\rr^j
\idirac0\rangle
\
=
\
\prod_{j=1}^{\infty}
\ll
\frac{1-t z^j}{1-z^j}
\rr^j 
\label{gf-multiplicative}
\end{eqnarray}

\noindent where we have used the fact that $\idirac \Gamma_{-}(z,t) = 
\idirac$, and $\Gamma_{+}(z,t)\fdirac = \fdirac$. 

\subsection{Equating the two results}

Equating the above results, we have a new derivation of the 
1-parameter extension of MacMahon's generating function, 
first obtained in \cite{v2}

\begin{equation}
\label{vuleticgf}
\sum_{\pi} A_{\pi}(t) z^{|\pi|} = \prod_{j=1}^{\infty}
\ll
\frac{1-t z^j}{1-z^j}
\rr^j
\end{equation}   

\subsection{Weighted plane partitions with restricted levels} 

{}From the form of $A_{\pi}(t)$ in (\ref{a-weight}), it is clear
that by setting the deformation parameter to a primitive $n$-th
root of unity, $t = e^{2\pi i/n}$, $n\in \N \geq 2$, $\mathcal{S}(t)$ 
in (\ref{gf-additive}--\ref{gf-multiplicative}) is the generating 
function of weighted plane partitions with paths of maximally level 
$(n-1)$.

\section{Connection with the KP integrable hierarchy}
\label{TF}

\subsection{Finite scalar products}
We consider $\mathcal{S}_{s\times s}(t)$, the generating function of 
plane partitions of the same type as those 
generated by $\mathcal{S}(t)$, but now the plane partitions are 
confined to an $s\times s$ 
square base (with no conditions on the column heights). 
$\mathcal{S}_{s\times s}(t)$ is obtained from the finite scalar 
product

\begin{eqnarray}
\label{finite-scalar-product}
\mathcal{S}_{s\times s}(t)
=
\idirac
\lprod_{j=1}^{s}
\Gamma_{+}(z^{\frac{-2j+1}{2}}, t)
\rprod_{k=1}^{s}
\Gamma_{-}(z^{\frac{2k-1}{2}}, t)
\fdirac
\end{eqnarray}

Next, we introduce dependence on two finite sets of parameters 
$\{u_1, \ldots, u_s\}$ and 
$\{v_1, \ldots, v_s\}$ by considering the more general finite 
scalar product 

\begin{equation}
\begin{split}
\label{GD-1}
\quad
&\mathcal{S}_{s\times s}(u_1, \ldots , u_s, v_1, \ldots , v_s; t)
=
\\
&\langle 0| 
\Gamma_{+}(u_{s}^{-1},t) \ldots \Gamma_{+}(u_1^{-1},t) 
\Gamma_{-}(v_1,t) \ldots \Gamma_{-}(v_s,t) 
|0 \rangle
\end{split}
\end{equation}

\noindent Evaluating the right hand side of (\ref{GD-1}) using (\ref{h-v3}), 
we obtain

\begin{equation}
\label{GD-2}
\mathcal{S}_{s\times s}(u_1, \ldots , u_s, v_1, \ldots , v_s; t)
=
\prod_{i,j=1}^{s}
\frac{1-t u_i v_j}{1- u_i v_j} 
\end{equation}

We note that $\mathcal{S}_{s\times s}(u_1, \ldots , u_s, v_1,
\ldots , v_s; t)$ 
is the restriction of a function of two infinite sets of variables 
${\bf u} = \{u_1, u_2, \ldots\}$ and 
${\bf v} = \{v_1, v_2, \ldots\}$, obtained by setting 
$u_t = v_t = 0$, for all $t > s$, and we show that 
this function is a KP $\tau$-function. 
More precisely, we show that there is an extension 
$A({\bf u}, {\bf v}; t)$ 
of 
$\mathcal{S}_{s\times s}(u_1, \ldots , u_s,$ $ v_1, \ldots , v_s; t)$ 
that is a $\tau$-function of KP in the power sum variables 
$x_m = \frac{1}{m} \sum_{j=1}^{\infty} u_j^m$. 

\subsection{An extension that is a non-trivial KP $\tau$-function} 

From chapter III, equation (4.7) in \cite{mac} 

\begin{equation}
\label{form-1}
\mathcal{S}_{s\times s}(u_1, \ldots , u_s, 
              v_1, \ldots , v_s; t)
=
\sum_{\{ \lambda | \  l(\lambda) \leq s \}}
s_{\lambda}(u_1, \ldots, u_s)
S_{\lambda}(v_1, \ldots, v_s; t)
\end{equation}

\noindent where $s_{\lambda}(u_1, \ldots, u_s)$ is a Schur
function
\begin{eqnarray}
s_\lambda(u_1,\ldots,u_s)
&=&
{\rm det} \ll h_{\lambda_i - i + j}(u_1,\ldots,u_s) \rr
\\
\sum_{m=0}^{\infty} h_m(u_1,\ldots,u_s) z^m
&=&
\prod_{j=1}^{s}  \frac{1}{(1-u_j z)} 
\end{eqnarray}

\noindent and 
\begin{eqnarray}
S_{\lambda}(v_1, \ldots, v_s; t) 
&=&
{\rm det} \ll q_{\lambda_i - i + j} (v_1,\ldots, v_s; t)\rr
\label{S}
\\
\sum_{m=0}^{\infty} q_m(v_1, \ldots, v_s; t) z^m
&=& 
\prod_{j = 1}^{s} \ll \frac{1- t v_j z}{1- v_j z} \rr
\label{q}
\end{eqnarray}

Now consider

\begin{equation}
\label{a-form2}
A({\bf u}, {\bf v}; t) = 
\sum_{\{ \lambda | \  l(\lambda) \leq s \}}
s_{\lambda}(u_1, u_2, \ldots)
S_{\lambda}(v_1, v_2, \ldots; t)
\end{equation}

\noindent where the sets 
$\{u_1, u_2, \ldots\}$ 
and  
$\{v_1, v_2, \ldots\}$
are now infinite, but the sum over the partitions $\lambda$ 
is still restricted, $\{ \lambda | \  l(\lambda) \leq s \}$. 
$\mathcal{S}_{s\times s}(u_1, \ldots ,$$ u_s, v_1, \ldots ,$$ v_s; $ $t)$
is the restriction of $A({\bf u}, {\bf v}; t)$ obtained by 
setting $u_t=v_t=0$ for all $t>s$. 
Letting $\frac{1}{m}
\sum_{j=1}^{\infty} u_j^m = x_m$ and 
$\frac{1}{m}\sum_{j=1}^{\infty} v_j^m = y_m$, 
we can write

\begin{equation}
\label{a-form}
A({\bf u}, {\bf v}; t) =
\sum_{\{ \lambda | \  l(\lambda) \leq s \}}
\chi_{\lambda}(x_1, x_2, \ldots)
\chi_{\lambda}(\tilde{y}_1, \tilde{y}_2, \ldots)
\end{equation}

\noindent where $\tilde{y}_m = (1 - t^m) y_m$, and 
$\chi_{\lambda}(x_1, x_2, \ldots)$ is the character 
polynomial

\begin{eqnarray}
\chi_\lambda(x_1,x_2,\ldots)
&=&
{\rm det} \ll p_{\lambda_i - i + j}(x_1, x_2, \ldots) \rr
\\
\sum_{m=0}^{\infty} p_m(x_1,x_2,\ldots) z^m
&=&
\exp \ll \sum_{m=1}^{\infty} x_m z^m \rr
\end{eqnarray}

The coefficients $\chi_{\lambda}(\tilde{y}_1, \tilde{y}_2, \ldots)$ 
are of the right form to satisfy the Pl\"ucker relations
non-trivially \cite{mjd}, hence $A({\bf u}, {\bf v}; t)$ 
is a non-trivial KP $\tau$-function in the variables 
$\{x_1, x_2, \ldots\}$ and 
$\mathcal{S}_{s\times s}(u_1, \ldots , u_s, v_1, \ldots , v_s; t)$ 
is the restriction obtained by setting $u_t=v_t=0$ for all $t>s$, 
in the power sums 
$\{x_1, x_2, \ldots\}$ and
$\{y_1, y_2, \ldots\}$.

\subsection{Another extension that is a trivial KP $\tau$-function} 
It is possible to write 
$\mathcal{S}_{s \times s} (u_1, \ldots,$$u_s,v_1,\ldots,$$v_s;t)$ 
as a restriction of another function $B({\bf u},{\bf v})$, as 
follows. 

\begin{eqnarray}
\nonumber
\mathcal{S}_{s \times s} (u_1, \ldots,u_s,v_1,\ldots,v_s;t)
&=&
\prod_{i,j=1}^{s}
\frac{1-t u_i v_j}{1- u_i v_j} 
\\
&=&
\nonumber
\exp
\ll
\sum_{i,j=1}^{s}
\ll\log(1-t u_i v_j)-\log(1-u_i v_j)\rr
\rr
\\
&=&
\label{}
\exp
\ll
\sum_{k=1}^{\infty}
\sum_{i,j=1}^{s}
\frac{(1-t^k)(u_i v_j)^k}{k}
\rr
\end{eqnarray}

Now consider

\begin{equation}
B({\bf u}, {\bf v}) 
= 
\exp
\ll
\sum_{k=1}^{\infty}
\sum_{i,j=1}^{\infty}
\frac{(1-t^k)(u_i v_j)^k}{k}
\rr
\end{equation}

\noindent where the sets 
$\{u_1, u_2, \ldots\}$
and
$\{v_1, v_2, \ldots\}$
are now infinite.
$\mathcal{S}_{s\times s}(u_1, \ldots ,$$ u_s, v_1, \ldots ,$$
v_s; $ $t)$
is the restriction of $B({\bf u}, {\bf v}; t)$ obtained by
setting $u_t=v_t=0$ for all $t>s$.
Letting $\frac{1}{k} \sum_{i=1}^{\infty} u_i^k = x_k$ and 
$\frac{1}{k}\sum_{j=1}^{\infty} v_j^k = y_k$, we can write

\begin{equation}
B({\bf u}, {\bf v}) 
= 
\exp
\ll
\sum_{k=1}^{\infty}
k(1-t^k)x_k y_k
\rr
\end{equation}

\noindent which is a trivial KP $\tau$-function, since it is 
the exponential of a linear function in $\{x_1,x_2,\ldots\}$, 
so it satisfies any nonlinear partial differential equation 
in the KP hierarchy trivially.

We conclude that it is possible to write 
$\mathcal{S}_{s \times s} (u_1, \ldots,u_s,v_1,\ldots,v_s;t)$ 
as a restriction of either 
$A({\bf u},{\bf v})$ or 
$B({\bf u},{\bf v})$. 
The advantage of the former is that it offers a connection 
with a non-trivial solution of KP, in contrast with the latter
which satisfies KP trivially.

\section{$(t, q)$-Operators and Macdonald weighted plane 
partitions}
\label{M}

\subsection{$(t, q)$-Fermions}

The $t$-fermions are a special case of more general 
$(t, q)$-fermions that depend on two deformation parameters 
$t$ and $q$. We obtain the latter as the components of the 
fermion generating functions

\begin{eqnarray}
\sum_{j \in \integer}
\psi_j z^j
&=&
\exp 
\ll 
\phantom{-}
\sum_{m=1}^{\infty}
\frac{1-t^m}{m (1-q^m)} h_{-m} z^m
\rr
\\
&\times&
\exp 
\ll
-\sum_{m=1}^{\infty}
\frac{1-t^m}{m (1-q^m)} h_{m} z^{-m}
\rr
\nonumber
\\
\sum_{j \in \integer}
\psis_j z^j
&=&
\exp 
\ll
-\sum_{m=1}^{\infty}
\frac{1-t^m}{m (1-q^m)} h_{-m} z^m
\rr
\\
&\times&
\exp 
\ll
\phantom{-}
\sum_{m=1}^{\infty}
\frac{1-t^m}{m (1-q^m)} h_{m} z^{-m}
\rr
\nonumber
\end{eqnarray}

\noindent where the $(t, q)$-Heisenberg generators $h_m$ satisfy

\begin{equation}
[h_m, h_n] = m \frac{1-q^{|m|}}{1-t^{|m|}} \delta_{m+n,0}
\end{equation}

The $t$-fermions and all related operators are obtained by 
setting $q=0$. {}From the $(t, q)$-fermion generating functions, 
one deduces the $(t, q)$-(half-)vertex operators

\begin{eqnarray}
\Gamma_{+}(z,t,q) &=& \exp 
\ll
-\sum_{m=1}^{\infty} \frac{1-t^m}{m (1-q^m)} h_m z^{-m} 
\rr
\\
\Gamma_{-}(z,t,q) &=& \exp 
\ll
-\sum_{m=1}^{\infty} \frac{1-t^m}{m (1-q^m)} h_{-m} z^m 
\rr
\end{eqnarray}

\noindent which satisfy the commutation relation

\begin{eqnarray}
\Gamma_{+}(z,t,q) \Gamma_{-}(z',t,q) 
&=& 
\prod_{n=0}^{\infty} \frac{1- q^n \frac{t z'}{z}}{1- q^n \frac{z'}{z}} 
\Gamma_{-}(z',t,q) \Gamma_{+}(z,t,q)
\label{macdonaldcom}
\\
&=&
\frac{(\frac{t z'}{z},q)_\infty}{(\frac{z'}{z},q)_\infty} 
\Gamma_{-}(z',t,q) \Gamma_{+}(z,t,q)
\nonumber
\end{eqnarray}

\subsection{Generating function for Macdonald weighted plane partitions}

Consider the scalar product

\begin{eqnarray}
\quad
\mathcal{S}(t,q)
&=&
\idirac \ldots 
\Gamma_{+}(z^{-\frac{3}{2}},t,q) \Gamma_{+}(z^{-\frac{1}{2}},t,q)
\Gamma_{-}(z^{\frac{1}{2}},t,q) \Gamma_{-}(z^{\frac{3}{2}},t,q)
\ldots \fdirac
\\
\nonumber
\\
&=&
\idirac
\lprod_{j=1}^{\infty}
\Gamma_{+}(z^{\frac{-2j+1}{2}},t,q)
\rprod_{k=1}^{\infty}
\Gamma_{-}(z^{\frac{2k-1}{2}},t,q)
\fdirac
\nonumber
\end{eqnarray}

Using (\ref{macdonaldcom}), one easily obtains 

\begin{equation}
\mathcal{S}(t,q) = \prod_{n=1}^{\infty}
\ll
\frac{(tz^n,q)_\infty}{(z^n,q)_\infty}
\rr^n
\end{equation}

\noindent which is the right-hand-side of the following equation 
for the generating function of 2-parameter weighted plane partitions 
derived combinatorially  in \cite{v2}

\begin{equation}
\label{vulmac}
\sum_{\pi} 
F_{\pi}(q,t) z^{|\pi|}
=
\prod_{n=1}^{\infty}
\ll
\frac{(tz^n,q)_\infty}{(z^n,q)_\infty}
\rr^n
\end{equation}

What is missing is a Fock space derivation of the left-hand-side 
of (\ref{vulmac}) which requires a proof of the following statements

\begin{eqnarray}
\label{macconditions}
\Gamma_{-}(z,t,q)|\mu\rangle & = & 
\sum_{\lambda \succ \mu} P_{\lambda \slash \mu} (z,t,q) 
|\lambda\rangle 
\\
\langle \mu |\Gamma_{+}(z,t,q) & = & 
\sum_ {\lambda \succ \mu} 
P_{\lambda \slash \mu} (z^{-1},t,q) 
\nonumber
\langle\lambda|
\end{eqnarray}

\noindent where $P_{\lambda\slash\mu}(z,t,q)$ is the skew Macdonald 
polynomial of a single variable $z$ \cite{mac}.

\section{Comments}
\label{C}

We obtained a Fock space derivation of the 1-parameter 
$t$-deformation of MacMahon's plane partition generating 
function in \cite{v2}, starting from the underlying $t$-fermions 
of \cite{j1}.
For $t = e^{2\pi i / n}$, we obtained generating 
functions of weighted plane partitions with horizontal paths that 
are maximally $n-1$ rim hooks wide. When formulated on a finite 
$s\times s$ square base, we showed that these generating
functions are evaluations of KP $\tau$-functions.

Our proofs are based on the action of underlying fermions on 
vector states. This complicates the analysis, because we start 
from the anti-commutation relations and we derive the action 
of vertex operators (which are exponentials in bilinears in the 
fermions) on Fock space state vectors. It is possible 
to bypass these complications and work entirely in terms of the 
action of the Heisenberg generators as in \cite{lam}. This 
simplifies the treatment, as one starts by requiring the 
vertex operator actions that one wants, such as 
(\ref{macconditions}) at the expense of losing contact with 
the underlying fermions.

In \cite{sul}, a connection between the $t$-deformation of 
MacMahon's generating function was discussed also on the 
basis of Jing's $t$-fermions. There it was pointed out, amongst 
other results, that for $t \in \R$ the expression (\ref{vuleticgf}) 
is the topological string partition function on a conifold and 
a geometric interpretation of the parameter $t$ (called $Q$ in 
\cite{sul}) was obtained. The derivation of (\ref{vuleticgf}) 
in \cite{sul} was based on commuting the vertex operators in 
the scalar product. Further, $t$-deformed free fermions and 
results that are related to those discussed in this work (and 
in \cite{sul}) were also obtained in \cite{tsilevich}. 

\section{Acknowledgements}
OF wishes to thank I~Grojnowski for a discussion that prompted
the start of this work, P~Bouwknegt for hospitality at ANU, 
Canberra, where it was started, A~Ram, K~Takasaki, J-Y~Thibon 
and M~Vuleti\'c for discussions on this and related topics. 
MW is supported by an Australian Postgraduate Award.

\end{document}